\documentclass[aps,prd,preprint]{revtex4-1}
\usepackage{graphicx,color}
\usepackage{amsmath,amssymb}
\usepackage{multirow}
\newcommand{\lsim}{\mathrel{\mathop{\kern 0pt \rlap
  {\raise.2ex\hbox{$<$}}}
  \lower.9ex\hbox{\kern-.190em $\sim$}}}
\newcommand{\gsim}{\mathrel{\mathop{\kern 0pt \rlap
  {\raise.2ex\hbox{$>$}}}
  \lower.9ex\hbox{\kern-.190em $\sim$}}}
\newcommand{\pb}{{\,{\rm pb}}}

\newcommand{\kev}{{\,{\rm keV}}}

\newcommand{\gev}{{\,{\rm GeV}}}
\newcommand{\tev}{{\,{\rm TeV}}}
\newcommand{\tb}{\bar{t}}
\newcommand{\al}{{\alpha}}

\newcommand{\sg}{{\sigma}}
\newcommand{\dt}{{\delta}}
\newcommand{\tht}{{\theta}}
\newcommand{\Dt}{{\Delta}}

\newcommand{\lm}{{\lambda}}
\newcommand{\Lm}{{\Lambda}}
\newcommand{\gm}{{\gamma}}

\newcommand{\Gm}{{\Gamma}}
\newcommand{\rd}{{\partial}}
\newcommand{\beq}{\begin{equation}}
\newcommand{\eeq}{\end{equation}}
\newcommand{\bea}     {\begin{eqnarray}}
\newcommand{\eea}     {\end{eqnarray}}
\newcommand{\no}{{\nonumber}}

\newcommand{\gmf}{{\gamma^5}}
\newcommand{\br}{{\rm Br}}

\newcommand{\Lg}{{\mathcal{L}}}

\newcommand{\n}{ {(n)} }
\newcommand{\on}{ {(1)} }
\newcommand{\tw}{ {(2)} }
\newcommand{\z}{ {(0)} }

\newcommand{\htw}{ h^{(2)} }
\newcommand{\bon}{  B^{(1)} }
\newcommand{\ton}{  t^{(1)} }
\newcommand{\bton}{  \bar{t}^{(1)} }
\newcommand{\phitw}{ \Phi^{(2)} }
\newcommand{\inty}{\frac{1}{2}\int_{-\pi R}^{\pi R}d y}
\newcommand{\dsl}{ \rlap{\,/}{D}}
\newcommand{\sut}{$SU(2)$}

\newcommand{\key}[1]{}

\begin{document}

\title{The second Kaluza-Klein neutral Higgs bosons \\
in the minimal Universal Extra Dimension model}
\bigskip
\author{Sanghyeon Chang}
\email{sang.chang@gmail.com}
\author{Kang Young Lee}
\email{kylee14214@gmail.com}
\author{Jeonghyeon Song}%
\email{jhsong@konkuk.ac.kr}
\affiliation{
Division of Quantum Phases \& Devises, School of Physics, 
Konkuk University,
Seoul 143-701, Korea
}
\date{\today}

\begin{abstract}
Loop-induced decay of a neutral Higgs boson into
a pair of gluons or photons has great implications
for the Higgs discovery at the LHC.
If the Higgs boson is heavy with mass above $\sim 500\gev$,
however, these radiative branching ratios are very suppressed
in the standard model (SM), 
as the new decay channels are kinematically open.
We note that these radiative decays can be sizable 
for the heavy CP-odd second Kaluza-Klein (KK) 
mode of the Higgs boson, $\chi^\tw$, in the
minimal universal extra dimension model:
highly degenerate mass spectrum of the theory
prohibits kinematically the dominant KK-number-conserving decays
into the first KK modes of the $W$, $Z$ and top quark.
We find that the CP-even decay of
$h^{(2)} \to g g$ is absent at one-loop level 
since $h^{(2)}$ couples with different mass eigenstates of $t_{1,2}^\on$
while a gluon does with the same mass eigenstates.
The $h^\tw$ production at the LHC is very suppressed.
On the contrary, 
the process $ gg \to \chi^\tw \to \gamma\gamma$
in an optimal scenario can be observed with
manageable SM backgrounds
at the LHC.
\end{abstract}

\maketitle

\section{Introduction}
\label{sec:introduction}

The universal extra dimension (UED) 
model\,\cite{ued}
has recently drawn a lot of interest as it suggests
solutions for proton decay\,\cite{proton},
the number of fermion generations\,\cite{fermion-generation-number},
and supersymmetry breaking\,\cite{susy-break}.
Based on a flat five-dimensional (5D)  spacetime,
this model assumes that all the standard model (SM) 
fields propagate in
the additional extra dimension $y$ with size $R$,
compactified over an $S_1/Z_2$ orbifold.
This {\it universal} accessibility
to the extra dimension protects
the Kaluza-Klein (KK) number conservation at tree level and
the KK parity conservation at loop level.
This new parity invariance has two significant implications
in the phenomenology.
First, the compactification scale 
can come down as low as about 300 GeV
since the contributions of the KK modes to electroweak 
precision observables
arise only through loops.
Second, the exact invariance of the KK parity
allows the cold dark matter candidate,
the lightest KK particle (LKP)\,\cite{KKcdm}.

The identity of the LKP depends crucially 
on the radiative corrections
to the KK masses.
There are two types of radiative corrections
to the KK mass.
The first is the bulk correction from compactification
over finite distances, which is
well-defined and finite.
The second type corrections are from the boundary kinetic terms,
which are incalculable due to unknown physics 
at the cutoff scale $\Lm$.
The minimal version of this model, called the mUED model, is based
on the assumption that the boundary kinetic terms vanish 
at the cut-off scale.
Then radiative corrections to the KK masses are
well-defined, leading to the first KK mode of the 
$U(1)_Y$ gauge boson $\bon$ 
as the LKP\,\cite{rad-correction}. 
Many interesting phenomenological signatures of the mUED
have been studied\,\cite{pheno}.

New particle contents and their phenomenology of the UED model
resemble those of a supersymmetry model
with $R$ parity conservation:
all the SM particles have their heavy partner with odd parity;
the decay of each heavy partner ends up with
the lightest new particle (missing energy signal)
plus some SM particles.
There are three distinctive features of, especially,
the mUED model.
First, the new heavy partner has the same spin 
as the corresponding SM particle.
In the literature,
the spin discrimination in supersymmetry 
and UED models have been studied extensively,
although very challenging at the LHC\,\cite{discrimination,spin}.
The second characteristic is
nearly degenerate mass spectra
of new particles\,\cite{bosonic-susy,discrimination}.
High degeneracy in the KK masses at the same KK level
makes the decay products of a heavy new particle consist of
very soft SM particles with missing energy.
At the LHC,
this is to be overwhelmed by QCD backgrounds.
The third characteristic is the presence of
\emph{even} KK parity heavy particles, the second KK modes.
The even parity allows their decay into
two SM particles,
which can be smoking-gun signatures of this model.
In Ref.~\cite{discrimination},
it was shown that 100 fb$^{-1}$
data of the LHC can discover the second KK modes of the $Z$ 
boson and the photon 
through the decays into two leptons.

In this paper, we focus on the massive scalar particles
with even KK parity,
the second KK modes of the Higgs boson.
The $n$-th KK modes of a \sut\  doublet Higgs field
consist of CP-even neutral $h^{(n)}$, 
CP-odd neutral $\chi^{(n)}$,
and charged scalars $\phi^{\pm (n)}$.
In the literature, the Higgs sector in the mUED model 
has been studied, mostly focused on 
the effects of the first KK modes.
The zero mode of the Higgs boson has 
${\mathcal O}(10\%)$ increase in its 
gluon fusion production 
and the ${\mathcal O}(10\%)$ decrease
in the $h \to \gamma\gamma$ decay width,
by the first KK mode effects through loops\,\cite{petriello}.
The phenomenological signature of 
$h^\on$ was also discussed,
concluding that the production
at the LHC is suppressed because the dominant channel is 
through the production and 
subsequent decay of the first KK mode of
the $b$ quark\,\cite{KKhiggs-kundu}.    
The detection of $h^\on$
is expected even more challenging 
because the decay products
involve too soft SM particles.
However the \emph{second}  KK Higgs bosons
can avoid these difficulties.

In this paper
we restrict ourselves to two neutral KK Higgs bosons, 
$\phitw = \htw,\chi^\tw$.
Kaluza-Kelin number conserving decay modes are
tree level decays of $\phitw$ into
$\bon\bon$, $\bon\chi^\on$, $\ton\bar{t}^\on$,
$W^\on W^\on$, and $Z^\on Z^\on$.
In the minimal model\,\cite{minimal-ued},
however, larger radiative contributions to the KK masses
of the top quark and \sut\ gauge bosons
prohibit these decays especially when the SM Higgs boson mass
is light around 120 GeV.
Even though tree level decay modes of
$\htw \to \bon\bon, \ell^{+(1)} \ell^{-(1)} $
and $\chi^{(2)} \to \ell^{+(1)} \ell^{-(1)} $ are still kinematically allowed,
the branching ratios are suppressed
because of very small phase space from nearly degenerate masses  
and/or small lepton Yukawa couplings.

\begin{figure}[t!]
\centering
\includegraphics[width=\textwidth]{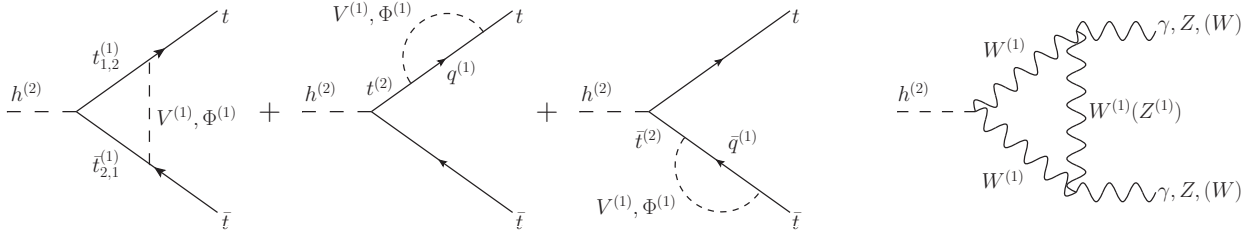}
\caption{\label{fig:Feyn:even}\small
The Feynman diagrams for the  decay of 
the CP-even  neutral Higgs bosons at one-loop level.
$\htw \to gg$ channel is
prohibited in this model.
$V^{(1)}$ are first KK modes of gauge bosons and
$\Phi^{(1)}=h^{(1)},\chi^{(1)}$.
}
\end{figure}
\begin{figure}[t!]
\centering
\includegraphics[width=\textwidth]{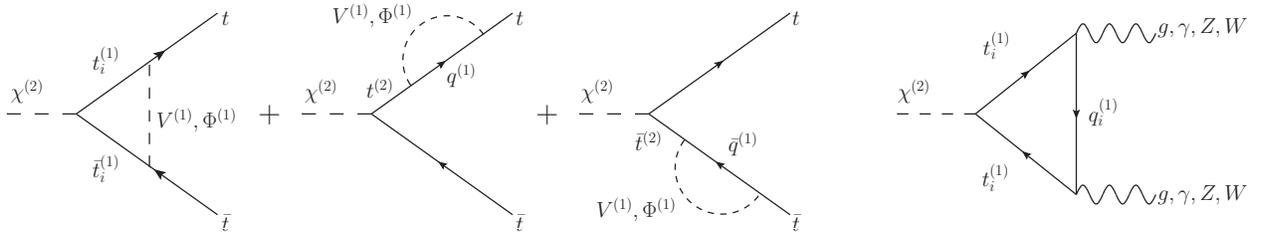}
\caption{\label{fig:Feyn:odd}\small
The Feynman diagrams for the decay of 
the CP-odd neutral Higgs bosons at one-loop level.
}
\end{figure}

In the mUED model, therefore, 
the loop-induced decay of the 
second KK Higgs boson 
can be substantial.
Their Feynman diagrams 
are illustrated in Figs.~\ref{fig:Feyn:even} and \ref{fig:Feyn:odd}.
As shall be shown,
the vertex of $h^{(2)}$-$g$-$g$
vanishes even at one-loop level because 
the 
CP-even $h^{(2)}$ couples with different
mass eigenstates of $t_{1,2}^\on$
while the gluon with the same mass eigenstates.
The production of the CP-even neutral Higgs boson
through gluon fusion at the LHC is very suppressed.
On the contrary, 
the decay of
the CP-odd scalar $\chi^\tw$ into
$gg$ is substantial, 
which leads to sizable gluon fusion production
at the LHC.
In addition, ${\rm BR}(\chi^\tw \to \gm\gm)$ is not extremely
small as in the SM.
At the LHC, the heavy CP-odd neutral Higgs boson
is produced through the gluon fusion,
and can be detected by two hard photons.
The SM backgrounds are to be shown manageable.
This is our main results.

The organization of the paper is as follows.
In the next section, we briefly review the model and describe
the effective interactions focused on the Higgs 
and top quark sector.
Section \ref{sec:lhc} deals with the production and decay 
of the second KK modes of the neutral Higgs bosons.
For the process of $gg \to \chi^\tw \to \gm\gm$,
we give details of the SM backgrounds and
the kinematic cuts to see the signal in Sec.~\ref{sec:back}.
We conclude in Sec.~\ref{sec:conclusions}.

\section{Brief review of the mUED model}
\label{sec:review}

The UED model is based on a flat 5D spacetime with the metric of
\bea
g_{MN} = 
\left(
\begin{array}{cc}
g_{\mu\nu} & 0 \\
0 & -1 \\
\end{array}
\right),
\eea
where $M,N=0,1,\cdots,4$,
and $g_{\mu\nu} = {\rm diag} (1,-1,-1,-1)$
is the 4D metric.
The 5D gamma matrices are $\Gm^M = (\gm^\mu, i \gm^5)$.
The size of the extra dimension $y$ is $R$.

In this model, all the SM fields
propagate freely in the 5D bulk. 
The zero mode of each 5D field corresponds to the 
SM particle.
In order to obtain chiral zero mode of a fermion
from a 5D vector-like fermion field,
we compactify the extra dimension on an $S^1/Z_2$ orbifold.
We assign odd parity under the $Z_2$ orbifold symmetry
to the zero mode fermion with wrong chirality.
This extends the fermion sector
to accommodate both $SU(2)$-doublet 
and $SU(2)$-singlet
SM fermions.
For the third generation quarks, \textit{e.g.},
we have
\bea
\label{eq:QTt}
Q_3(x,y) = 
\left(
\begin{array}{c}
T(x,y) \\
B(x,y)\\
\end{array}
\right),
\quad
t(x,y),
\quad b(x,y).
\eea
In addition, the fifth-dimensional gauge field $V_5(x,y)$
has odd $Z_2$ parity.

Focused on the phenomenology of the second KK modes of
the Higgs boson, 
we present the KK expansion of a gauge boson $V_M(x,y)$,
the Higgs field $H(x,y)$, the $SU(2)$-doublet top quark
$T(x,y)$, and the $SU(2)$-singlet top quark
$t(x,y)$:
\begin{eqnarray}
V_\mu(x,y)
&=&
\frac{1}{\sqrt{\pi R}}
\left[
V^\z_{\mu}(x)
 + \sqrt{2} \sum_{n=1}^\infty V^\n_\mu (x) \cos \frac{n y}{R}
\right],
\\ \no
V_5(x,y)
&=&
\sqrt{\frac{2}{\pi R}}
\sum_{n=1}^\infty V^\n_5 (x) \sin \frac{n y}{R},
\\ \no
H(x,y) &=& \frac{1}{\sqrt{\pi R}}
\left[
H^\z(x) + \sqrt{2} \sum_{n=1}^\infty H^\n(x) \cos \frac{n y}{R}
\right],
\\ \no
T(x,y) &=&
\frac{1}{\sqrt{\pi R}}
\left[
T^\z_{L}(x) + \sqrt{2} \sum_{n=1}^\infty 
\left\{
T_{L}^\n (x) \cos \frac{n y}{R}
+ 
T_{R}^\n (x) \sin \frac{n y}{R}
\right\}
\right],
\\ \no
t(x,y) &=&
\frac{1}{\sqrt{\pi R}}
\left[
t^\z_R(x) + \sqrt{2} \sum_{n=1}^\infty 
\left\{
t_{R}^\n (x) \cos \frac{n y}{R}
+ 
t_L^\n (x) \sin \frac{n y}{R}
\right\}
\right],
\end{eqnarray}
where $V^M =B^M, W^M, A^M$
and $f_{R,L} = (1\pm \gm^5)f/2$ for a fermion $f$.
Here $n$ is called the KK number. 
Note that, \textit{e.g.}, the KK modes of $SU(2)$-doublet top quark
have both chiralities.

At tree level, the KK mass is
\bea
\label{eq:tree:KK:mass}
M_{KK}^{(n)} = \sqrt{  M_n^2 +  m_0^2},
\eea
where $M_n=n/R$, and $m_0$ is the corresponding
SM particle mass.
All the KK mode masses are highly degenerate.
In Ref.~\cite{rad-correction},
it was shown that the radiative corrections 
generate significant changes in the KK masses.
In the minimal model
based on the assumption of vanishing boundary 
kinetic terms at the cutoff scale $\Lm$,
the corrections are well-defined and finite.

\subsection{The Higgs sector}
\label{subsec:higgs}
The 4D effective Lagrangian in the Higgs sector is
obtained by integrating out the extra dimensions $y$:
\begin{eqnarray}
\label{eq:Lg:Higgs}
\Lg_{H} = \inty
\left[
(D_M H )^\dagger D^M  H
+\mu^2  H^\dagger  H
-\frac{\lm_{h5}}{2} (H^\dagger  H)^2
\right],
\end{eqnarray}
where $D_M$ is the covariant derivative given by
$D_M H = (\rd_M -\frac{ i}{2} g_5 \tau^iW^i_M 
- \frac{i}{2} g'_5 B_M )$.
The 4D SM Higgs boson interaction is recovered if
$g^{(\prime)} = g^{(\prime)}_5/\sqrt{\pi R}$
and $\lm_h = \lm_{h5} /\sqrt{\pi R}$.
The Lagrangian in Eq.~(\ref{eq:Lg:Higgs}) yields
the following 4D potential of the Higgs boson:
\bea
V_{\rm eff}
&=& - \mu^2 H^{\z \dagger} H^\z 
+ \frac{\lm_h}{2} \left( H^{\z \dagger} H^\z  \right)^2
\\ \no
&&
+ \sum_{n=1}^\infty \left( M_n^2 - \mu^2 \right)
H^{\n \dagger} H^\n
+ \frac{1}{4} \lm_h \sum_{n,m,\ell,k=1}^\infty
H^{(n)\dagger} H^{(m)} H^{(\ell)\dagger}H^{(k)}
\Dt^2_{n,m,\ell,k},
\eea
where 
\bea
\Dt^2_{n,m,\ell,k} = 
\dt_{k,\ell+n+m}
+\dt_{\ell,n+m+k}
+\dt_{n,m+k+\ell}
+ \dt_{m,n+k+\ell}
+\dt_{k+m,n+\ell}
+\dt_{k+n,m+\ell}
+\dt_{k+\ell,m+n}.
\eea
Positive $\mu^2$ (or negative mass squared)
generates
non-zero vacuum expectation value (VEV) for 
the SM Higgs boson $H^\z$,
which triggers the electroweak symmetric breaking.
However the condition $R^{-1} > \mu$ leads to
positive mass squared parameters for all the KK Higgs bosons:
the KK Higgs bosons do not have non-zero VEV.

The $n$-th KK mode of the $SU(2)$-doublet Higgs boson is
\bea
H^\n (x) = 
\left(
\begin{array}{c}
\phi^{\n +}
\\
\frac{1}{\sqrt{2}} 
\left(
h^\n + i \chi^\n
\right)
\end{array}
\right),
\eea
where $h^\n$ and $\chi^\n$ are the CP-even 
and CP-odd neutral scalar fields, respectively.
The mass eigenstate of CP-odd scalar 
$\chi_Z^\n$ is a linear combination
of $\chi^\n$ and the fifth component of the $n$-th KK mode
of the $Z$ boson, $Z^{5 \n}$:
\bea
\label{eq:chi:mixing}
\chi_Z^\n = \frac{M_n \chi^\n + m_Z Z^{5 \n} }{\sqrt{M_n^2 + m_Z^2}}
\equiv \cos\theta_\chi^\n \chi^\n + \sin\theta_\chi^\n Z^{5\n}.
\eea
Its orthogonal combination is the Goldstone mode $G_Z^\n$ for
the $Z_\mu^\n$\,\cite{petriello}.
Note that $\chi_Z^\n$ and $Z^\n$ have the same mass at tree level. 
The KK masses of neutral Higgs bosons are
\bea
\label{eq:KK:CP:even:Higgs:mass}
m_{h^\n}^2&=& M_n^2 + m_h^2 +\delta m_{H^\n}^2,
\\
\label{eq:KK:CP:odd:Higgs:mass}
m_{\chi^\n}^2&=& M_n^2 + m_Z^2 +\delta m_{H^\n}^2,
\eea
where $m_h^2=\lm_h v^2$
and $v \approx 246\gev$ is the VEV of the SM Higgs boson. The radiative mass correction 
to $n$-th KK scalar masses is 
\bea
\label{eq:rad:Higgs:mass}
\delta m_{H^\n}^{ 2}=M_n^2\left(\frac{3}{2}g^2+\frac{3}{4}{g'}^2
-\lm_h \right)\frac{1}{16\pi^2}\ln\frac{\Lambda^2}{\mu^2},
\eea
where $\mu$ is the regularization scale\,\cite{rad-correction}.
For the second KK mode production, 
we put $\mu=2R^{-1}$\,\cite{KKhiggs-kundu}.

\subsection{The top quark sector}
\label{subsec:top}
Due to large top quark mass, 
there is non-negligible mixing between the KK modes of
\sut-doublet and \sut-singlet top quarks in the same KK level.
Their 4D effective Lagrangian is
\bea
\label{eq:Lg:top}
\Lg_{t}
= \inty 
\left[
i \bar{T} \dsl T
+ i \bar{t} \dsl t
- \left(\lm_{t5} \bar{Q}_3 \tilde{H} t + H.c. \right)
\right],
\eea
where $\tilde{H} = i \sg_2 H^*$.
As the zero mode of the Higgs boson develops non-zero
VEV of $v$,
the mass matrix of the KK top quark becomes
non-diagonal.
Including the radiative corrections 
to the mass,
the $n$-th KK mass term for the top quark is
\bea
-{\mathcal L}_{\rm mass} = \sum_{n=1}^\infty
\left(\bar T^\n_L, \bar t^\n_L\right)
\left(
\begin{array}{cc}
M_n + \dt m_{T^\n} & m_t \\
m_t & -M_n - \dt m_{t^\n}
\end{array}
\right)
\left(\begin{array}{c}
       T^\n_R \\ t^\n_R
      \end{array}
\right)
,
\eea
where $\dt m_{T^\n}$ and $\dt m_{t^\n}$
are, respectively, the radiative corrections to the \sut-doublet and 
\sut-singlet top quarks,
given by \,\cite{rad-correction}
\bea
\dt m_{T^\n} &=&
\frac{M_n}{16\pi^2} \left(
3 g_s^2 + \frac{27}{16} g^2
+ \frac{1}{16} g^{\prime 2}
- \frac{3 }{4} y_{t}^2 
\right) \ln\,\frac{\Lm^2}{\mu^2},
\\ \no
\dt m_{t^\n} &=&
\frac{M_n}{16\pi^2} \left(
3 g_s^2
+  g^{\prime 2}
- \frac{3 }{2} y_{t}^2 
\right) \ln\,\frac{\Lm^2}{\mu^2},
\eea
where $y_t = m_t/v$.
The KK mass of
the $SU(2)$-doublet top quark $T^\n$ has larger radiative correstions
than that of the $SU(2)$-singlet top quark $t^\n$.

Two mass eigenstates of the $n$-th KK mode
are denoted by $t_1^\n$ and $t_2^\n$.
Here $t_1^\n$ is the lighter mass eigenstate. 
The mass eigenstates are related with electroweak eigenstates $t^\n$
and $T^\n$ through the mixing angle $\tht_t^\n$:
\bea
\left(
\begin{array}{c}
t^\n_{1R} \\ t^\n_{2R}
\end{array}
\right)
&=&
\left(
\begin{array}{rr}
~\cos \frac{\tht_t^\n}{2} & ~-\sin  \frac{\tht_t^\n}{2} \\
~\sin  \frac{\tht_t^\n}{2} &~\cos \frac{\tht_t^\n}{2} 
\end{array}
\right)
\left(
\begin{array}{c}
t^\n_{R} \\T^\n_{R} 
\end{array}
\right),
\\ \no
\left(
\begin{array}{c}
t^\n_{1L} \\ t^\n_{2L}
\end{array}
\right)
&=&
\left(
\begin{array}{rr}
\cos \frac{\tht_t^\n}{2} & ~-\sin  \frac{\tht_t^\n}{2} \\
-\sin  \frac{\tht_t^\n}{2} & ~-\cos \frac{\tht_t^\n}{2} 
\end{array}
\right)
\left(
\begin{array}{c}
t^\n_{L} \\T^\n_{L} 
\end{array}
\right).
\eea
The mixing angle $\tht_t^\n$ is
\bea
\tan\tht_t^\n = \dfrac{m_t}{ M_n 
+ \dfrac{\dt m_{T^\n}+\dt m_{t^\n} }{2}},
\eea
and the physical masses are, to a good approximation,
\bea
m_{t^\n_1} = 
\sqrt{ \left( M_n + \dt m_{t^\n} \right)^2+m_t^2 },
\quad
m_{t^\n_2} = 
\sqrt{ \left(M_n + \dt m_{T^\n} \right)^2+m_t^2 }.
\eea

\section{The LHC reach for the second KK Higgs bosons}
\label{sec:lhc}

The 4D 
interaction Lagrangian for $\htw$ is
\begin{eqnarray}
\label{eq:Lg:int}
\Lg_{\htw} &=& 
-y_{t(2)} h^\tw \bar{t}t
+i y_{t(2)} \chi^\tw \bar{t}\gm_5 t
- y_t \htw \left( \tb_1^\on t_2^\on + 
\tb_2^\on t_1^\on \right)
\\ \no &&
+\frac{g^{\prime 2}v}{4}  \htw 
\bon_\mu B^{\on \mu}
+
\frac{g^2 v}{2\sqrt{2}}h^\tw W^{\on\dagger}_\mu W^{\on\mu}
+
\frac{g^2 v}{2\sqrt{2}}h^\tw Z^{\on\dagger}_\mu Z^{\on\mu} 
\\ \no &&
+
\frac{g'}{2\sqrt{2}}
\left[
(\partial^\mu \chi^\on ) h^\tw -\chi^\on \partial^\mu h^\tw
\right]B_\mu^\on,
\end{eqnarray}
and that for $\chi_Z^\tw$ is
\begin{eqnarray}
\label{eq:Lg:int:chi2}
\Lg_{\chi_Z^\tw} &=& 
- i y_t  \cos\theta_\chi^\tw \chi_Z^\tw 
\left[
\sin\tht_t^\on
\left(
\tb_1^\on  \gmf t_1^\on + \tb_2^\on \gmf t_2^\on
\right)
+
\cos\tht_t^\on
\left( \tb_1^\on t_2^\on -
\tb_2^\on t_1^\on \right)
\right]
\\ \no
&& - i \frac{g}{\sqrt{2}} \sin\theta_\chi^\tw  \chi_Z^\tw 
\tb_2^\on  \gm_5 t_2^\on
\\ \no 
&&- e Q_t
A_\mu^\z 
\left[
\bton_1 \gm^\mu \ton_1
+
\bton_2 \gm^\mu \ton_2
\right]
- g_s
\left[
\bar{t}^{\on a}_1 \gm^\mu {\bf G}_{\mu}^{\z } t^{\on b}_1
+
\bar{t}^{\on a}_2 \gm^\mu {\bf G}_{\mu}^{\z }t^{\on b}_2
\right]
,
\end{eqnarray}
where ${\bf G}_{\mu}^{\z } \equiv  G_{\mu}^{\z c}T^c_{ab}$
and $a,b,c$ are the color indices.
We have shown interactions to leading order in the 
small KK mixing angle $\theta_{t,\chi}^\n$.
Note that the second line of Eq.~(\ref{eq:Lg:int:chi2})
is from the interaction of $Z^5$ through the mixing
in Eq.~(\ref{eq:chi:mixing}).
Since the second KK mixing angle $\theta_\chi^\tw$
is smaller than the first KK mixing angle $\theta_\chi^\on$,
the effect of the second line of Eq.~(\ref{eq:Lg:int:chi2})
is subleading.
The vertex of $h^\tw \bar{t}t$ at one-loop level
is \cite{Kakizaki}
\bea
\label{eq:yt2}
y_{t(2)} = 
\frac{y_{t}}{48\sqrt{2}\pi^2}
\left(
16 g_s^2 - \frac{39}{4}g^2 +\frac{4}{3} g^{\prime 2}
- 9 y_{t}^2 + 3 \lm_h
\right) \ln \frac{\Lm}{\mu}.
\eea
Similar expressions for other KK fermions such as
the KK tau lepton can be inferred 
with the replacement of $\theta_\tau \sim m_\tau R \ll 1$.

The tree level decay rates of CP-even $\htw$ 
are
\begin{eqnarray}
\label{eq:H2:tree:decay:VV}
\Gamma ( \htw \to V^\on V^\on)
&=& \mathcal{S}\frac{g_{H_2 V_1 V_1}^2}{64\pi}
 \frac{1}{m_{h^{(2)}}}
\frac{1-4x_V^2+12 x_V^4}{x_V^4} \sqrt{1-4x_V^2},
\\ \label{eq:H2:tree:decay:chiB} 
\Gamma ( \htw \to \chi^\on \bon)
&=&
\frac{g^{\prime 2}}{128 \pi} 
\frac{m_{\htw}^3}{m_{B^\on}^2}
\lm^{3/2}\left( x^2_{\chi^\on}, x^2_{B^\on}\right),
\\ \label{eq:H2:tree:decay:tt}
\Gamma ( \htw \to t^\on_1\bar{t}^\on_2)
&=&
N_C \, y_{t}^2 \frac{m_{h^{(2)}}}{16 \pi}
\left[1-\left(x_{t_1}+x_{t_2}\right)\right]^{3/2}
\left[1-\left(x_{t_1}-x_{t_2}\right)\right]^{1/2},
\eea
where $V^\on_\mu = B_\mu^\on, W_\mu^\on, Z_\mu^\on$, 
$\mathcal{S}$ is the symmetric factor 
($\mathcal{S}=1/2$ for $V^\on_\mu = B_\mu^\on, Z_\mu^\on$ and
$\mathcal{S}=1$ for $V^\on_\mu=W_\mu^\on$),
$x_i = m_{i}/m_{h^{(2)}}$,
and $\lm(a,b)=1+a^2+b^2-2a-2b-2ab$.
The vertices of $\htw$-$V^\on$-$V^\on$ are
\bea
g_{H_2 B_1 B_1} = \frac{g^{\prime 2}}{2} v,
\quad
g_{H_2 Z_1 Z_1} = g_{H_2 W_1 W_1} = \frac{g^{ 2}}{2\sqrt{2}} v.
\eea
Note that $g_{H_2 Z_1 Z_1}$ and
$g_{H_2 W_1 W_1}$ are the same to leading order
because of very small KK Weinberg angle~\cite{rad-correction}.
Compared to the SM coupling of $h$-$W$-$W$,
$g_{H_2 W_1 W_1}$ has additional factor of $1/\sqrt{2}$.

At one-loop level, 
$\htw$ decays into
a pair of top quarks, and a pair of photons.
Their decay rates are
\begin{eqnarray}
\label{eq:H2:decay}
\Gamma ( \htw \to t\bar{t})
&=&
N_C \, y_{t(2)}^2 \frac{m_{h^{(2)}}}{8 \pi}
(1-4x_t^2)^{3/2},
\\ \no
\Gamma ( \htw \to \gm\gm)
&=&
\frac{g^2\alpha^2 m_W^2}{128 \pi^3 m_{h^{(2)}}}
\left|
\hat{{\mathcal A}}_1^H(\tau_{W^\on})
\right|^2 
,
\eea
where $\tau_i = 1/(4 x_i^2)= m_{h^{(2)}}^2/4m_i^2$.
The normalized amplitude for spin-1 particles is given by 
\bea
\hat{{\mathcal A}}_1^H(\tau)  =
     -\left[ 2 \tau^2+3 \tau + 3(2\tau-1)f(\tau) \right]\tau^{-1},
\eea
where the universal scalar function $f(\tau)$ is
\bea
\label{eq:ftau}
f(\tau) = \left\{
\begin{array}{l l}
\arcsin ^2 \sqrt{\tau} & \hbox{ if } \tau \leq 1,\\
-\dfrac{1}{4}
\left[
\ln \dfrac{1+\sqrt{1-\tau^{-1}}}{1-\sqrt{1-\tau^{-1}}} - i \pi
\right]^2 & \hbox{ if } \tau>1.\\
\end{array}
\right.
\eea

The CP-odd $\chi^\tw$ has only radiative decays
because of its small mass both in the light and heavy $m_h$ cases.
The decay rates are
\bea
\label{eq:chi2:decay}
\Gamma ( \chi^\tw \to t\bar{t})
&=&
N_C \, y_{t(2)}^2 \frac{m_{\chi^{(2)}}}{8 \pi}
(1-4x_t^2)^{1/2},
\\ \no
\Gamma ( \chi^\tw \to \gm\gm)
&=&
\frac{G_F \al^2 m_{\chi^{(2)}}^3}{128 \sqrt{2}\pi^3}
\left(
\frac{m_t}{m_{t^\on}}
\right)^2
\left|
\sum_{i=\ton_1,\ton_2}
N_C Q_t^2 \sin\tht {\mathcal A}^A_{1/2}(\tau_{i})
\right|^2,
\\ \no
\Gamma ( \chi^\tw \to g g)
&=&
\frac{G_F \al_s^2 m_{\chi^{(2)}}^3}{36 \sqrt{2}\pi^3}
\left(
\frac{m_t}{m_{t^\on}}
\right)^2
\left|
\frac{3}{4} 
\sum_{i=\ton_1,\ton_2}
\sin\tht {\mathcal A}^A_{1/2}(\tau_i)
\right|^2.
\end{eqnarray}
The amplitude for spin-1/2 particles is
\bea
{\mathcal A}^A_{1/2}(\tau) &=& 2 \tau^{-1} f(\tau),
\eea
where $f(\tau)$ is given in Eq.~(\ref{eq:ftau}).
For more complicated expressions 
of $\Gamma(\htw/\chi^\tw \to WW,ZZ,Z\gamma)$
we refer the reader to Ref.~\cite{Djouadi,ZZ}.

Brief comments on the lower bounds on $R^{-1}$ are in order here.
Indirect observables put rather strong
constraint on $R^{-1}$.
Electroweak precision data 
with the subleading new physics contributions
and two-loop corrections to the SM $\rho$ parameter
leads to
$R^{-1} \gsim 600\, (300) \gev $ for $m_h = 115 \,(600)\gev$
at 90\% confidence level \cite{Gogoladze:2006br}. 
In addition, the $B\to X_s\gm$ branching ratio 
constrains this model more seriously
since the mUED KK modes interfere destructively
with the SM amplitude~\cite{Haisch:2007vb}.
At the 95\% (99\%) confidence level, the bound is
$R^{-1} \gsim 600 ~(300) \gev$.
Since we are focused on the direct probe of this model,
we take flexible parameter space of $R^{-1} \in [350,600]\gev$
as marginally allowed by indirect constraints,
which is commonly searched in the literature~\cite{recent}.
For the Higgs boson mass, we take two cases, the light
Higgs boson case of $m_h =120\gev$ and the heavy Higgs boson
case of $m_h=600\gev$.

At tree level, only the KK-number-conserving interactions
are possible:
a second KK mode mainly decays into two first KK mode particles.
The mass spectra of the first and second
KK modes determine the kinematic permission of 
each decay channel.
In Table \ref{table-uedmass},
we show the KK masses of the first KK modes of CP-odd Higgs boson,
gauge bosons, top
quark and tau lepton as well as the second KK modes
of the CP-even and CP-odd Higgs boson
in the mUED model. 
We set $\Lm R=20$, and take $R^{-1} =350,400,500\gev$
for $m_h = 120,\,600\gev$ cases.

\begin{table}[t!]
\caption{\label{table-uedmass}
The masses of KK states which can be involved in the second KK
state of the Higgs boson. 
We include one-loop radiative corrections and
fix $\Lambda R =20$ and $m_h =120\gev$. Masses and the decay rate
are in units of GeV.
}
\centering
\begin{ruledtabular}
\begin{tabular}{|cc|rrr|cccccc|}
~~$R^{-1}$& $m_h$ & $\htw$ & $\chi_Z^\tw$ & $\chi_Z^\on$ &
$\bon$ & $W^{\pm(1)}$ & $\ton_1$ & $\ton_2$ 
& $\tau^{\on}_1$ & $\tau^{\on}_2$\\
\hline
\multirow{2}{*}{350} & 120 & 
{715.9}  & {711.7}  & 365.3 &  
\multirow{2}{*}{351.4} & \multirow{2}{*}{377.5} &  
\multirow{2}{*}{410.9} & \multirow{2}{*}{428.4}
& \multirow{2}{*}{354.0} & \multirow{2}{*}{360.0} \\
 & 600 & 881.2 & 651.8 & 327.0 &  & &   & &  &  \\
\hline
\multirow{2}{*}{400} & 120 & 
{815.7}  & {811.8} &  414.5 &
\multirow{2}{*}{401.3} & \multirow{2}{*}{429.2} &  
\multirow{2}{*}{459.2} & \multirow{2}{*}{479.2} 
& \multirow{2}{*}{404.5} & \multirow{2}{*}{411.4} \\
 & 600 & 950.8 & 699.0 &  370.3 & & &   & &  &  \\
\hline
\multirow{2}{*}{500} & 120 & 
{1015.4}  & {1012.4} &  513.5 &
\multirow{2}{*}{500.9} & \multirow{2}{*}{533.2} &  
\multirow{2}{*}{558.1} & \multirow{2}{*}{583.0} 
& \multirow{2}{*}{505.7} & \multirow{2}{*}{514.2} \\
 & 600 & 1100.0 & 926.5 & 457.8 & & &   & &  &  \\
\end{tabular}
\end{ruledtabular} 
\end{table}

The SM Higgs boson mass affects most drastically
the KK mass of the Higgs boson.
As in Eq.~(\ref{eq:tree:KK:mass}),
the tree level KK mass consists of the geometrical mass
and the corresponding SM particle mass.
In the heavy SM Higgs boson case,
large $m_h$ makes 
the mass of $\htw$ much larger
than $M_2(=2R^{-1})$, as in Eq.~(\ref{eq:tree:KK:mass}):
the KK mass degeneracy is broken for $\htw$.
An interesting observation is that
large $m_h$
or large Higgs quartic coupling $\lm_h$
makes \emph{negative} contributions 
through the radiative corrections 
as in Eq.~(\ref{eq:rad:Higgs:mass}).
This negative $\delta m^2_{H^\n}$ contribution
applies to the CP-odd $\chi^\n$ identically,
while the tree level KK mass of $\chi^\n$ has 
the SM $Z$ boson mass, not the SM Higgs boson mass.
Therefore, the KK modes of the CP-odd Higgs boson
become lighter as $m_h$ increases.

In the light Higgs boson case ($m_h=120\gev$), 
tree level decays of the second KK Higgs bosons
are 
$h^\tw \to \bon \bon$, $h^\tw/\chi^\tw \to \ell^\on \ell^\on$,
and $h^\tw/\chi^\tw \to \ell^\tw \ell$.
All other decay channels are kinematically closed,
and  CP-\emph{odd} $\chi^\tw$ 
cannot decay into two $\bon$'s.
In addition, leptonic decay modes are numerically negligible.
Even the most dominant leptonic decay mode
has 
$\Gm(h^\tw \to \tau^\on \tau^\on) \sim $ keV,
which is very suppressed by small Yukawa coupling of the tau lepton
and the very limited kinematic phase space 
because of $m_{\tau^\on} \approx 0.5 m_{h^\tw}$.
Another tree level decay mode,
$h^\tw/\chi^\tw \to \ell^\tw_R \ell$, is also very suppressed
because of the same reasons.
For definiteness, we present the masses of the second 
KK tau leptons
and $\Gm( \htw \to \tau_R^\tw \tau)$
with the fixed $R\Lm =20$ and $m_h=120\gev$:
\bea
\begin{array}{c|ccc}
R^{-1} & m(\tau_R^\tw) & m(\tau_L^\tw) 
& \Gm(\htw \to \tau_R^\tw \tau) \\
\hline
~~~350\gev~~~ & ~~~706.1 \gev~~~ & ~~~715.4\gev~~~ & ~~~20.2\kev~~~ \\
400\gev & 807.0 \gev & 817.6\gev & 18.0\kev \\ 
500\gev & 1008.7 \gev & 1021.9\gev & 13.9\kev
\end{array}
\eea
Thus the dominant tree level decay mode of $h^\tw$
in the light Higgs boson case is into a pair of LKP's.
Since $B^\on$ is a CDM candidate in this model,
this decay does not leave any track in the detector,
and appears as a missing energy signal.

At one-loop level, the KK number conservation is broken
while the KK-parity is still preserved. 
Thus the second KK mode of the Higgs boson can decay 
into two SM particles through loops mediated by
first KK modes.
First $\htw$ and $\chi^\tw$ can decay into a pair of top quarks
through the triangle diagram
(see Figs.~\ref{fig:Feyn:even} and \ref{fig:Feyn:odd}).
As shall be shown, this $t\tb$ decay mode is
dominant for both $\htw$ and $\chi^\tw$.
Unfortunately the huge SM $t\tb$ backgrounds
obstruct the observation of the signal.
Second types of radiative decays  are
into a SM gauge boson pair of $WW$, $ZZ$, $Z\gm$, $\gm\gm$
and $gg$.
For the decay of $\htw$ and $\chi^\tw$
these decays  are through the first KK gauge boson 
and the first KK top quarks respectively.

The radiative decays of $\Phi^\tw(=\htw,\chi^\tw)$
into a pair of gluons or photons require more detailed discussion.
The decay $\htw\to gg$
is through
the loop mediated by
the first KK modes of the top quark, as illustrated
in Figs.\,\ref{fig:Feyn:even} and \ref{fig:Feyn:odd}.
However the effective vertex of 
$\htw$-$g$-$g$ at one-loop level
vanishes because the CP-even scalar $\htw$ does not interact with
two identical mass eigenstates of the first KK top quarks
as can be seen in Eq.~(\ref{eq:Lg:int}).
Since the photon and gluon couple with $t_i^\on t_i^\on$,
the decay of $\htw \to gg$
and thus the gluon fusion production of $\htw$ are not possible.
On the contrary, the CP-odd scalar $\chi^\tw$
couples with the same mass eigenstates of the KK top quarks, 
although the coupling strength is suppressed
by the KK top mixing angle of the order of $ m_t /M_1$.
The decay of $\chi^\tw \to gg$
and its gluon fusion production are feasible at the LHC.

If the SM Higgs boson is heavy, \textit{e.g.},
$m_h=600\gev$,
the second KK mode of CP-even Higgs boson becomes also
heavy as in Table \ref{table-uedmass}.
Now $\htw\to W^\on W^\on, Z^\on Z^\on$ decay mode is open.
Another interesting decay mode is into $\chi^\on B^\on$,
as suggested by the third line of Eq.~(\ref{eq:Lg:int}).
This mode is dominant since
the vertex of $\htw$-$\chi^\on$-$B^\on$ is 
proportional to the $\htw$ mass,
while that of $\htw$-$W^\on$-$W^\on$ is proportional
to the SM gauge boson mass.
Finally
$\htw\to t_{1,2}^\on \bar{t}_{2,1}^\on$ is also kinematically allowed.

\begin{figure}[h!]
\centering
\includegraphics[width=4.in]{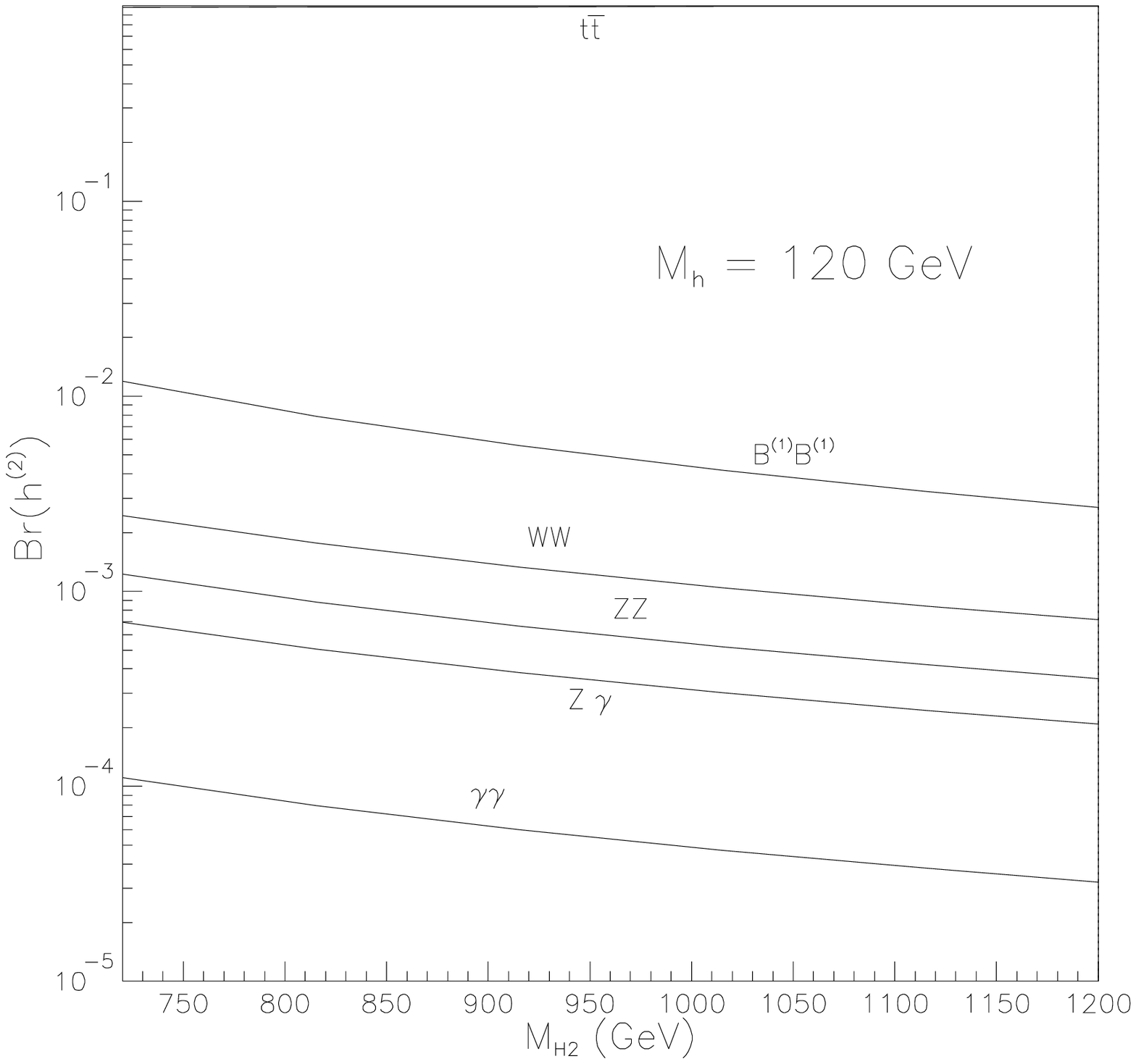}
\caption{\label{fig:br:H2:light:mh}\small
The branching ratios of the CP-even  
second KK neutral Higgs boson, $\htw$,
as functions of its mass.
We set $m_h=120\gev$ and $\Lambda R=20$.
}
\end{figure}

In Fig.~\ref{fig:br:H2:light:mh}, we present
the branching ratio of $\htw$ as a function of its mass
in the light SM Higgs boson case ($m_h=120\gev$).
The deacy into $\tb t$ is dominant because of the large
Yukawa coupling and strong coupling as in Eq.~(\ref{eq:yt2}).
The next dominant decay mode is KK-number conserving decay
of $\htw\to\bon\bon$.
This invisible branching ratio is about 1\%.
Narrow kinematic phase space from the degenerate mass spectrum
suppresses this decay.
Following decay modes are $WW$, $ZZ$, $Z\gm$, and $\gm\gm$.

\begin{figure}[t!]
\centering
\includegraphics[width=4.in]{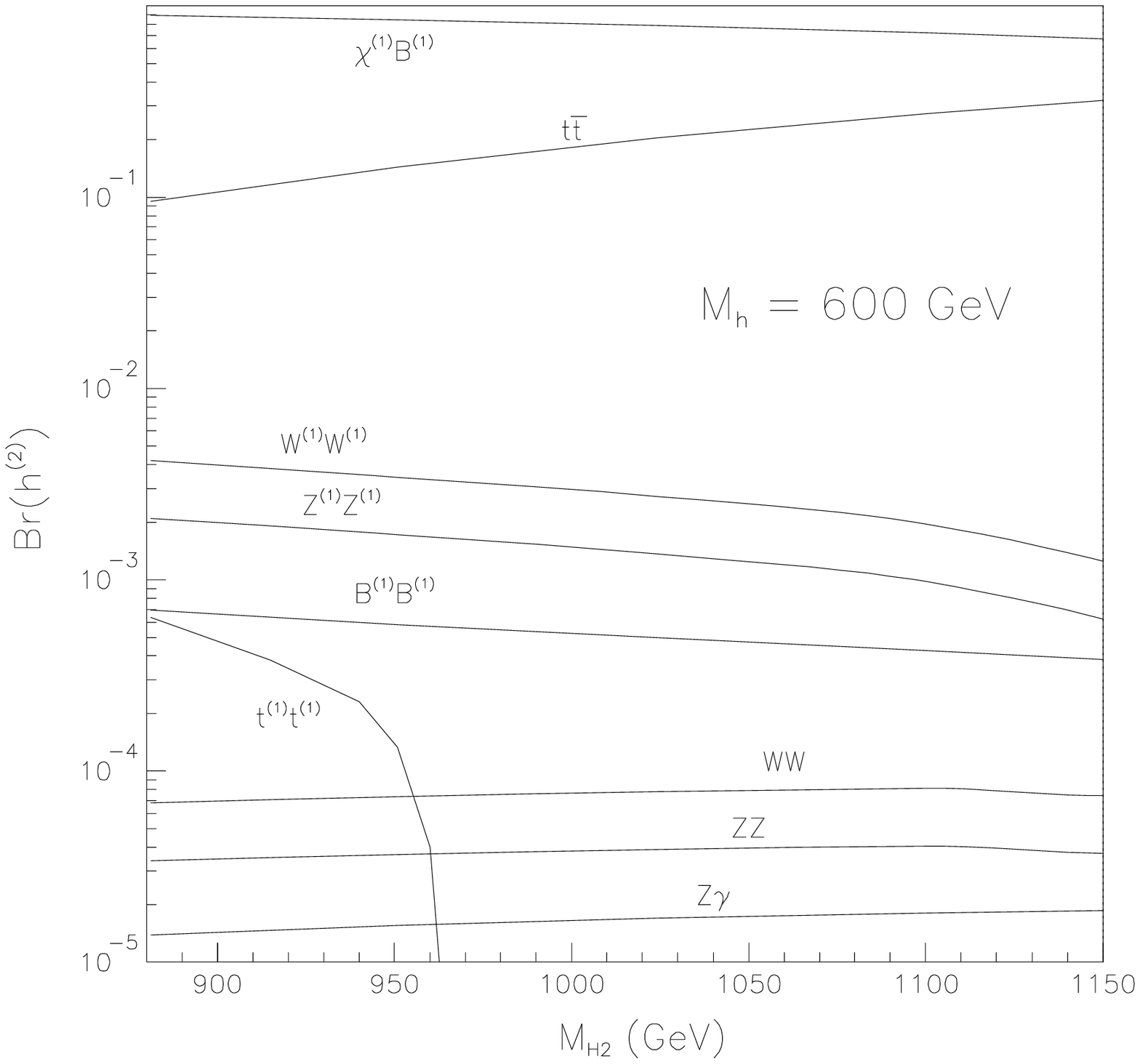}
\caption{\label{fig:br:H2:heavy:mh}\small
The branching ratios of the CP-even 
second KK neutral Higgs bosons, $\htw$,
as functions of its mass. We set $m_h=600\gev$ and $\Lambda R=20$.
}
\end{figure}

Figure \ref{fig:br:H2:heavy:mh} presents the branching
ratios of $\htw$ in the heavy SM Higgs boson case.
Large $m_h$ enhances
$\htw$ mass significantly.
In this case,
the decay into $\chi^\on B^\on$ is the most dominant one.
This can be understood by the large vertex $\htw$-$\chi^\on$-$\bon$
which is proportional to the heavy $\htw$ mass,
while the vertex $\htw$-$V^\on$-$V^\on$ is proportional to
the SM gauge boson $m_{W,Z}$.
The produced $\chi^\on B^\on$ decays through 
$\htw \to \chi^\on B^\on \to h^* \bon \bon
\to \bar{b} b \bon \bon$.
At the LHC, this signal is overwhelmed by QCD backgrounds.
The next dominant decay mode is
into a top quark pair, which becomes more significant
as $m_{\htw}$ increases.
Kaluza-Klein number conserving modes into
$W^\on W^\on$, $Z^\on Z^\on$, $\bon\bon$,
and $t^\on \tb^\on$ follow.
Note that for $R^{-1}>420\gev$ and $m_h=600\gev$,
$\htw\to t^\on \tb^\on$ is not kinematically allowed.
Radiative decays into a pair of SM gauge bosons 
are very suppressed.

\begin{figure}[t!]
\centering
\includegraphics[width=4.in]{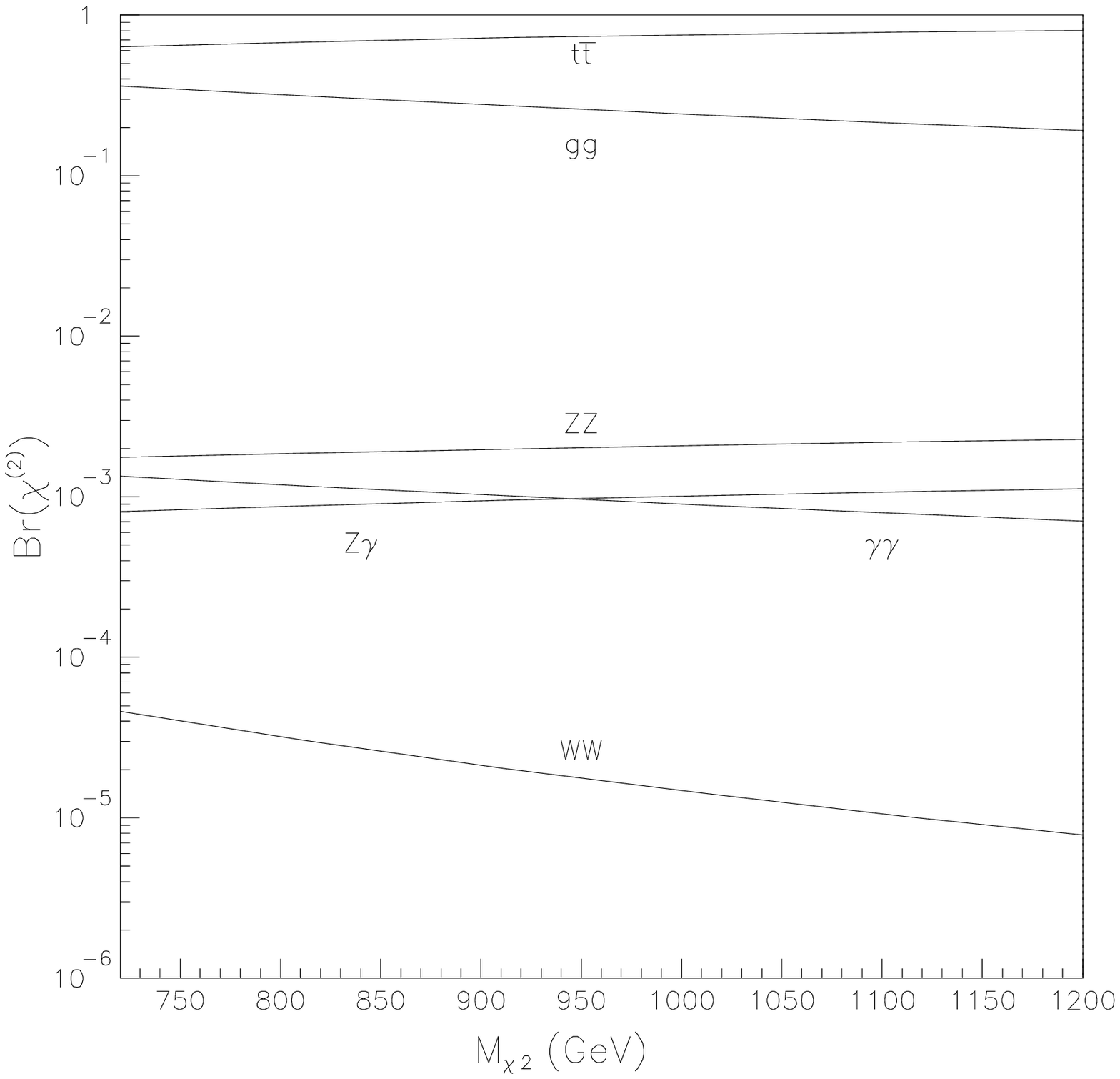}
\caption{\label{fig:br:A2:light:mh}\small
The branching ratios of the CP-odd 
second KK neutral Higgs bosons, $\chi^\tw$,
as functions of its mass. We set $m_h=120\gev$ and $\Lambda R=20$.
}
\end{figure}

The CP-odd Higgs boson $\chi^\tw$
does not have large enough mass for
KK-number-conserving decays.
Only radiative decays are allowed.
This pattern remains the same for the heavy Higgs boson case
because the $\chi^\tw$ mass decreases with increasing $m_h$
as discussed before.
In Fig.~\ref{fig:br:A2:light:mh}, 
we present the branching ratios of $\chi^\tw$
only in the light $m_h$ case.
The leading decay mode of $\chi^\tw$
is into a pair of top quarks.
The next dominent one is into a gluon pair with
${\rm BR}(\chi^\tw \to g g) \approx 20-40\%$.
We expect quite efficient production of $\chi^\tw$
through the gluon fusion at the LHC.
Decays into a pair of the SM gauge bosons
follow, in the order of $ZZ$, $\gm\gm$, $Z\gm$, and $WW$.
For the detection of $\chi^\tw$ at the LHC,
the $\gm\gm$ mode is expected to be most efficient.
The dominant decay mode into $t\tb$ suffers from large SM
background with the cross section of $\sim 900\pb$ \cite{top:factory}.
Other channels into $W$'s or $Z$'s
have additional suppression from their small branching ratios
of leptonic decay.
The decay into $\gm\gm$ has the branching ratio of 
$\sim 0.1\%$.
As shall be seen below,
$gg\to\chi^\tw\to\gm\gm$ in an optimal scenario
has a good chance to be observed 
at the LHC.

In all three cases of Figs.~\ref{fig:br:H2:light:mh}$-$\ref{fig:br:A2:light:mh},
$t\tb$ decay mode is dominant,
even though it is generated at one-loop level.
Suppressed KK-number-conserving decays 
are attributed to degenrate masses and thus small phase space.
Much smaller branching ratios of the decays into the SM gauge bosons
than that into $t\tb$ can be understood by two factors.
First, the coupling strength of $\htw$-$t$-$\tb$,
of which the dominant part is proportional to $y_t g_s^2$,
is much larger than that of
$\htw$-$V_\mu$-$V_\nu$ which is proportional to $ g^3$.
Second the decay amplitude
of $\htw\to t \tb$ is characterized by $m_{\htw}$
while that of $\htw\to V_\mu V^\mu$ by
$m_W$.

\begin{figure}[t!]
\centering
\includegraphics[width=4.2in]{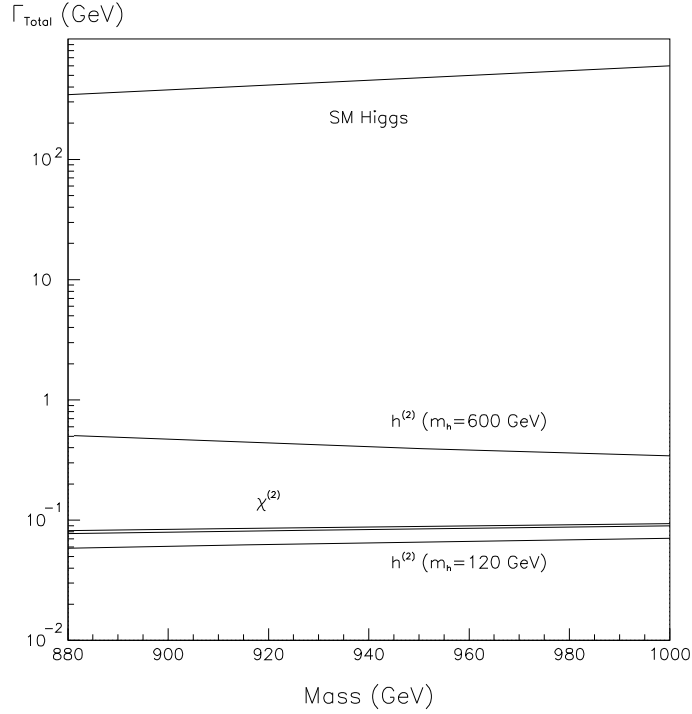}
\caption{\label{fig:totalwidth}\small
The total decay widths of $\chi^{(2)}$, $h^{(2)}$ and 
the SM Higgs boson with respect to their masses.
}
\end{figure}

In Fig. \ref{fig:totalwidth}
we compare the total decay widths of $\chi^{(2)}$
and $h^{(2)}$
with that of the SM Higgs boson.
We set $\Lm R =20$, and $m_h=120,600\gev$ for $h^{(2)}$.
The kinematic closure of many KK-number-conserving decays 
suppresses their total decay widths quite a lot.
The second KK Higgs bosons, even though very heavy, are
not obese like the SM one.
At a collider, they are expected to appear as resonances.

\begin{figure}[t!]
\centering
\includegraphics[width=4.2in]{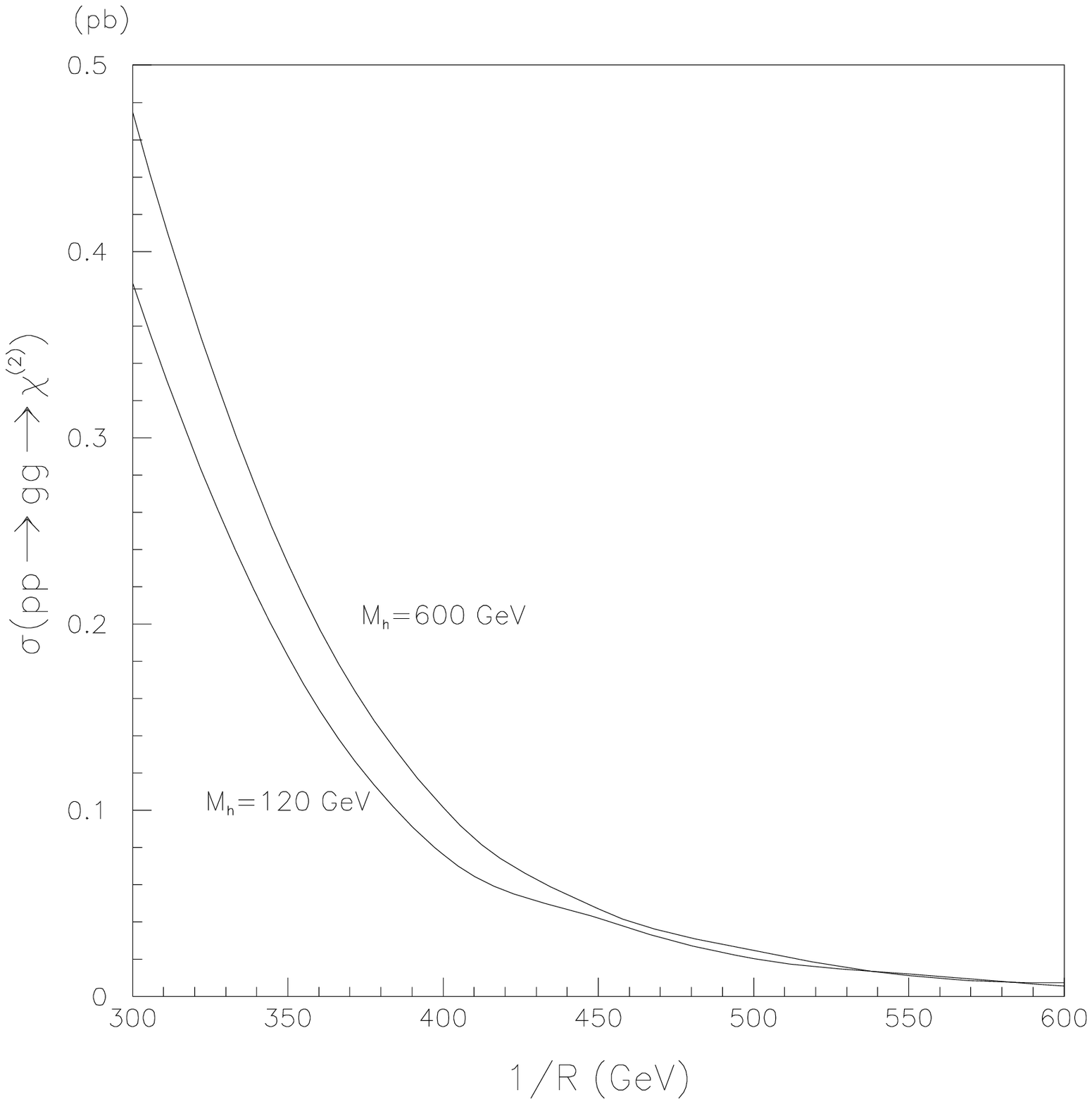}
\caption{\label{fig:production}\small
The cross section of $pp \to gg \to \chi^{(2)}$
at the LHC with respect to the mass of $\chi^{(2)}$.
}
\end{figure}

At the LHC, the most promising production is that
of $\chi^{(2)}$ through gluon fusion process,
$pp \to gg \to \chi^{(2)}$.
The production cross section at the parton level is given by
\beq
\hat{\sigma}(gg \to \chi^{(2)}) 
= \frac{4 \pi^2}{9 m_{\chi^{(2)}}^3}
\Gamma ( \chi^\tw \to g g).
\eeq
Figure \ref{fig:production} shows the production cross section
$\sigma(pp \to gg \to \chi^{(2)})$
as a function of $R^{-1}$ at the LHC
with $\sqrt{s}=14\tev$.
We take two cases of $m_h=120,600\gev$.
For the parton distribution function, we have used
the MRST 99\,\cite{mrst99}.
In the heavy SM Higgs boson case,
the production cross of $pp \to gg \to \chi^\tw$
is larger than that in the light Higgs case with the given $R^{-1}$.
It is mainly because of lighter $\chi^\tw$ mass with large
$m_h$, as shown in Table \ref{table-uedmass}.
In addition, large $m_h$ case is much less
constrained by indirect observables such as 
electroweak precision data and $B \to X_s \gm$:
for $m_h=600\gev$,
$R^{-1}>300\gev$ and $m_{\chi^\tw}>560\gev$.

Assuming the LHC integrated luminosity of 100 fb$^{-1}$, 
about 10,000 events of $\chi^{(2)}$ production are 
expected for $R^{-1}=500\gev$.
The most of $\chi^{(2)}$'s decay into a pair of top quark
or gluon jets,
which suffers from huge QCD backgrounds.
For heavier $\chi^\tw$ with mass above 1 TeV, 
top tagging becomes efficient and 
it can be a good channel to test the model.
For $R^{-1} <500$ GeV, however, 
top tagging efficient drops too 
much\,\cite{Kaplan:2008ie,Bhattacherjee:2010za}.
The next dominant decay modes
are into $ZZ$, $Z\gm$ and $\gm\gm$.
Considering small leptonic branching ratio of $Z$,
detection efficiency for the $Z$ boson 
is low.  
Thus $\chi^{(2)} \to \gamma \gamma$ is most promising decay channel 
to test the mUED model 
for $300$ GeV $\leq R^{-1} \leq 600$ GeV.
Since the branching ratio of 
${\rm BR}(\chi^{(2)} \to \gamma \gamma)$
is about $0.1\%$, we will have 
dozens of events of $\gamma \gamma$
pair from the $\chi^{(2)}$ decays.

For the optimal case of the detection of
$pp\to gg\to \chi^\tw \to \gm\gm$, 
we take 
\bea
m_h=600\gev,\quad R^{-1}=300\gev,\quad m_{\chi^\tw}=560\gev.
\eea
We adopt the K-factor of 1.3, which
represents the enhancement from higher order QCD processes
\cite{collider}.
Then the $\chi^\tw$ production cross section 
at the LHC is about 0.61 pb.
With $\br(\chi^\tw \to \gm\gm)\approx 10^{-3}$
and the integrated luminosity of 100 fb$^{-1}$,
$\sigma_{\chi^\tw} \br(\chi^\tw \to\gm\gm)$ has about 60 events.
And the invariant mass distributions of two photons
will show a resonant peak at the $\chi^\tw$ mass.
This special decay of $\chi^{(2)} \to \gamma \gamma$ 
can be a smoking gun signature to discriminate the mUED from SM or 
minimal supersymmetric standard model (MSSM) 
type heavy Higgs decay\,\cite{Djouadi}.

\section{Background Study}
\label{sec:back}

The photon events suffer from huge backgrounds from QCD processes.
Here, we estimate the backgrounds at the LHC 
with appropriate kinematic cuts
to check the significance of our signal.
In the ordinary $H \to \gamma \gamma$ analysis of the SM,
the background events are classified into two groups,
the irreducible backgrounds coming from two isolated photons 
at the parton level
and the reducible backgrounds including at least one fake photon.
Fake photons are mostly from the decays of $\pi^0$'s.
For the SM Higgs boson with mass about 150 GeV,
the two types of backgrounds 
are compatible to each other\,\cite{atlas}.  
For the heavy Higgs boson with mass $\gsim 500\gev$,
however, the irreducible backgrounds 
are negligible since their
subprocesses such as
$p p \to q \bar{q} \to \gamma \gamma$ and
$p p \to g g \to q \bar{q} \gamma \gamma$
decrease as $\sqrt{\hat{s}}$ increases.
On the contrary,
the reducible QCD backgrounds 
become relatively more important in this high energy region,
which are the main backgrounds of our two photon signal.

We calculate the cross sections 
for the dominant high $p_T$ QCD subprocesses 
by using PYTHIA\,\cite{pythia}.
We have applied the basic cuts 
of $p_T \ge 30$ GeV and $|\eta| \le 2.44$.
The dominant cross sections are
\bea
\sigma(g g \to g g )=6.91 \times 10^{8}~~{\rm pb},
\nonumber \\
\sigma(q g \to q g )= 8.71 \times 10^{8}~~{\rm pb},
\nonumber \\
\sigma(g g \to q \bar{q} )= 1.23 \times 10^{7}~~{\rm pb}.
\eea
Other subprocesses have much smaller cross sections. 
With integrated luminosity of 100 fb$^{-1}$, 
the total number of background events 
is as huge as $1.57 \times 10^{14}$.

In order to suppress the QCD background,
we first select the photons of 
$E_\gamma > 50$ GeV in the simulated events 
and take the most energetic two photons as the photon candidates.
Our kinematic cuts are chosen 
based on the characteristic features of 
the photons from the background events and the signal:
(i) most of the background photons are 
in the forward and backward directions along the beam line;
(ii) two photons of our signal events are 
in the opposite directions in the partonic c.m. frame.
Therefore, we apply 
kinematic cuts to exclude the forward and backward photons
along the beam line,
as well as the collinear photons.
The following kinematic cuts are called the \texttt{CUT I}:
\begin{description}
  \item[\texttt{CUT I} (1)] Transverse momentum cuts of $p_T > 30$ 
GeV for both photons are applied.
  \item[\texttt{CUT I} (2)] We demand that the opening angle 
of two photons are to be
$-1 < \cos \theta < -0.8$.
  \item[\texttt{CUT I} (3)] No other photons are collinear 
to the photon candidates,
where the collinear photon is defined by  
$0^\circ < \theta < 20^\circ$.
\end{description}
Having applied \texttt{CUT I}, 
we reduce the background events by five order of magnitude.

For the next step, we use the longitudinal boost invariance
at the LHC.
The two photons in our signal
have back-to-back momenta in the transverse plane.
We apply the following \texttt{CUT II}:
\begin{description}
  \item[\texttt{CUT II} (1)] The magnitudes of the transverse 
momenta of two photons
are same, 
\beq
-0.01 < \frac{p_{1T}-p_{2T}}{p_{1T}+p_{2T}} < 0.01.
\eeq
  \item[\texttt{CUT II} (2)] The opening angle of the transverse 
momenta of two photons are 
in the opposite direction, $-1 < \cos \theta_T < -0.985$.
\end{description}
With the \texttt{CUT II} applied, 
the background events are reduced
by three order of magnitude,
leaving $5.4 \times 10^{6}$ events
as the SM backgrounds.

Finally we apply the kinematic cut
on the invariant mass distribution of two photons.
Most QCD background photons have their invariant mass
distribution in the low mass region, less than 300 GeV.
With both \texttt{CUT I} and \texttt{CUT II}, 
there are less than 50 events per 10 GeV.
On the contrary, our signal $\chi^{(2)}\to \gm\gamma$
in the optimal scenario 
has about 61 events.
Therefore we have a very sharp peak 
over the SM backgrounds.

\section{Conclusions}
\label{sec:conclusions}
The probe of massive Higgs bosons 
with mass above $500\gev$
beyond the observed SM particles
is an interesting possibility at the LHC.
Within the SM, the Higgs boson
can be that heavy.
In the MSSM, additional heavy CP-even and CP-odd neutral 
Higgs bosons can be good candidates.
At the LHC, however, their detection is very challenging.
The production of this heavy Higgs boson,
mainly through the gluon fusion, is reduced by the kinematic suppression.
And the detection is not clean:
the SM heavy Higgs boson is too obese ($\Gamma_{h_{\rm SM}}
\simeq m_{h_{\rm SM}}$) to clearly declare the observation
from the golden $ZZ \to 4\ell$ mode;
in the decoupling limit
the MSSM heavy neutral Higgs bosons 
mainly decay into $t\tb$ and $b \bar{b}$
for the small and large $\tan\beta$ case, respectively,
which suffer from the QCD backgrounds. 

We found that the second KK modes of the
Higgs boson in the mUED model are also very interesting candidates
for massive Higgs bosons.
And they have very distinctive features
from the heavy Higgs bosons in the SM and MSSM.
Highly degenerate mass spectrum within the given KK level
closes kinematically most of the KK-number-conserving decays
into the first KK modes.
This kinematic closure leads to quite distinctive phenomenology
compared with the heavy Higgs boson(s) in the SM and MSSM.
First their total decay width is much small, 
which leads to a sharp resonance at the LHC.
The second characteristic
is the large branching ratio of CP-odd $\chi^\tw$ decay 
into two photons or two gluons.

It is also remarkable that
$\htw \to gg,\gm\gm$ through the KK fermion (mainly top) loops
is prohibited since the coupling
of the CP-even second KK Higgs boson
with the first KK fermions are off-diagonal.
The $\htw$ production through the gluon fusion 
is not feasible at the LHC.
On the contrary, the CP-odd $\chi^\tw$
has diagonal Yukawa couplings, though suppressed by
the factor of $m_t /M_1$.
Both $\chi^\tw \to gg$ and $\chi^\tw \to \gm\gm$
are allowed at one-loop level.
The CP-odd $\chi^\tw$ can be produced through the gluon fusion.
With the sizable ${\rm BR}(\chi^\tw\to\gm\gm)$,
the resonance in the $\gm\gm$ invariant mass distribution
gives a clear probe of the mUED model.

\acknowledgments
This work is supported by WCU program through the KOSEF funded
by the MEST (R31-2008-000-10057-0).
KYL is also supported by
the Basic Science Research Program 
through the National Research Foundation of Korea (NRF) 
funded by the Korean Ministry of
Education, Science and Technology (2009-0076208).
SC is supported  by the Basic Science Research Program 
through the NRF funded by the Korean Ministry of
Education, Science and Technology  (KRF-2008-359-C00011).
SC thanks KIAS 
for warm support during his visit.

\end{document}